\documentclass[aps,twocolumn,prl,showpacs]{revtex4}
\usepackage{epsfig}
\usepackage{isolatin1}

\usepackage{makeidx} \makeindex

\usepackage{amsmath} \usepackage{amssymb} \usepackage{amsthm}

\usepackage{mathrsfs} 

\begin{document}

\title {Compensation of decoherence from telegraph noise  by means of
bang-bang control}

\begin{abstract}
With the growing efforts in isolating solid-state qubits from
external decoherence sources, the origins of noise inherent to the
material start to play a relevant role. One representative example
are charged impurities in the device material or substrate, which
typically produce telegraph noise and can hence be modelled as
bistable fluctuators. In order to demonstrate the possibility of
the active suppression of the disturbance from a {\em single}
fluctuator, we theoretically implement an elementary bang-bang
control protocol. We numerically simulate the random walk of the
qubit state on the Bloch sphere with and without bang-bang
compensation by means of the stochastic Schr\"odinger equation and
compare it with an analytical saddle point solution of the
corresponding Langevin equation in the long-time limit. We find
that the deviation with respect to the noiseless case is
significantly reduced when bang-bang pulses are applied, being
scaled down approximately by the ratio of the bang-bang period to
the typical flipping time of the bistable fluctuation.  Our
analysis gives not only the effect of bang-bang control on the
variance of these deviations, but also their entire distribution.
As a result, we expect that bang-bang control works as a high-pass
filter on the spectrum of noise sources. This indicates how the
influence of $1/f$-noise ubiquitous to the solid state world can
be reduced.
\end{abstract}

\author{Henryk Gutmann} \affiliation{Sektion Physik and CeNS,
Ludwig-Maximilians-Universit\"at, 80333 M\"unchen, Germany}
\author{William M. Kaminsky} \affiliation{Department of Physics,
Massachusetts Institute of Technology, Cambridge, Massachusetts 02139}
\author{Seth Lloyd} \affiliation{Department of Mechanical Engineering,
Massachusetts Institute of Technology, Cambridge, Massachusetts 02139}
\author{Frank K. Wilhelm}  \affiliation{Sektion Physik and CeNS,
Ludwig-Maximilians-Universit\"at, 80333 M\"unchen, Germany}

\email[email: ]{gutmann@theorie.physik.uni-muenchen.de}


\maketitle

In order to implement solid-state quantum information processing
devices, the decoherence acting on the quantum states has to be
carefully understood, controlled and eliminated. So far, research
has concentrated on decoupling from external noise sources (like
thermal heat baths and electromagnetic noise). With the success of
this effort, noise sources intrinsic to the material such as
defect states increase in importance and have to be controlled in
order to improve coherence even further.

Most external noise sources are composed of extended modes in the
thermodynamic limit close to equilibrium such that their
fluctuations are purely Gaussian. Thus, their influence can be
modelled by an oscillator bath, see e.g. \cite{WeissQDS}.
However, there are physical situations when this assumption fails
\cite{GMB02,Stamp,PFFF02}. In particular, this is true for localized
noise sources with bounded spectra as they occur in disordered
systems for hopping defect states \cite{D&H}. Physical examples
for this situation are background charges in charge qubits
\cite{PFFF02,Ensslin,Zorin} or traps in the oxide layers of
Josephson tunnel junctions \cite{Harlingen}. Such localized noise
sources are more realistically represented by a collection of
bistable fluctuators \cite{PFFF02,TIYT03} (henceforth abbreviated
bfls), as their noise spectrum is considerably non-Gaussian. If
many of these noise sources with different flipping times are
appropriately superimposed, they lead to $1/f$ noise
\cite{D&H,Weissmann,Martinis}. With the progress of fabrication technology
and miniaturization of qubits, we expect however that there might
only be a few fluctuators in a qubit \cite{Harlingen}.

We analyze the impact of a single fluctuator in the semiclassical
limit, where it acts as a source of telegraph noise. We apply an
open loop quantum control technique, namely quantum bang-bang
\cite{L&V98,LVK99a,LVK99b}, which is designed suitably for slowly
fluctuating noise sources. We simulate the noise-influenced qubit
dynamics with and without bang-bang correction by integrating the
time-dependent Schr\"odinger equation for each specific
realization of the noise. We present the resulting random walks
around the unperturbed signal on the Bloch sphere and analyze the
quality of this suppression by an comparison of the ensemble
averaged deviations of these random walks with and without
bang-bang correction.

We describe our system by the effective Hamiltonian
\begin{eqnarray}
H_q^{\rm eff} (t) & = & H_{\rm q} + H_{\rm q,bfl}^{\rm noise}(t) \\
H_{\rm q} = \hbar \epsilon_{\rm q} \hat{\sigma}_{\rm z}^{\rm q} +
\hbar \Delta_{\rm q} \hat{\sigma}_{\rm x}^{\rm q} &\ & H_{\rm q,
bfl}^{\rm noise}(t) = \hbar \alpha \xi_{\rm bfl}(t)
\hat{\sigma}_{\rm z}^{\rm q} \label{eq:Hamiltonian}
\end{eqnarray}
where $\alpha$ denotes the coupling strength between the
fluctuator and the qubit and $\xi_{\rm bfl}(t)$ represents a
symmetric telegraph process that is flipping between $\pm 1$,
whose switching events are Poisson distributed with a mean
separation $\tau_{\rm bfl}$ between two flips.

On a microscopic level, such noise is typically generated by coupling 
the qubit to a two-state impurity, which is in turn coupled
to a heat bath causing the two-state system to flip randomly and
incoherently.  Our model corresponds to the semiclassical limit
and should be accurate whenever the coupling of the
impurity to the bath is much stronger than its coupling to the
qubit \cite{PFFF02,GMB02} such that the qubit does not act back on
the noise source. The assumption of a {\em symmetrical} telegraph
process corresponds to a high bath temperature compared to the
impurity level spacing. This restriction is not essential for the
following investigations, for an asymmetric noise signal would
only produce an additional constant drift.

We describe the resulting evolution of the noise influenced qubit
by a stochastic Schr\"odinger equation \cite{vKp,Arn73} with the
time-dependent Hamiltonian (\ref{eq:Hamiltonian}). For any initial
state of the qubit, we numerically integrate
\begin{equation}
\label{ssesolution}
\psi(t)  =  {\rm T} \exp \left({-i}/{\hbar} \int_0^t H_{\rm q}^{\rm
eff} (t^\prime)\,dt^\prime  \right) \psi(0)
\end{equation}
with ${\rm T}$ the time-ordering operator and $\psi$ the state
vector of the two-state system. The result is a random walk on the
Bloch sphere, which is centered around the free precession corresponding 
to $\epsilon_{\rm q}$ and $\Delta_{\rm q}$.

We implement the following idealized open quantum control scheme:
apply an infinite train of $\pi$-pulses on the qubit with
negligibly short pulse durations and a constant separation time
$\tau_{\rm bb}$ between neighboring pulses.  In doing so, we
intend to average out the $\hat{\sigma}_z$ parts of the effective
Hamiltonian (and thereby in particular the noise term) on time
scales large compared to $\tau_{\rm bb}$. This is accomplished by
iteratively spin-flipping the qubit and thus effectively switching
the sign of the noisy part of the Hamiltonian. This mechanism thus
works analogously to the well-known spin-echo procedure,
specifically the Carr-Purcell procedure of NMR \cite{CarrPurcell}.
We expect to compensate a fraction of the telegraph noise effects:
the size of the random walk induced by the noise is determined by
the typical time separation of the fluctuators influence between
two flips $\tau_{\rm bfl}$ and its coupling strength $\alpha$ and
scales roughly with $\alpha\tau_{\rm bfl}$ \cite{PFFF02}. Using bang-bang, 
the bfls influence remains uncompensated for at most a single
bang-bang period. Thus, we
reduce  the influence of the bfl randomly by an average factor
of $\tau_{\rm bfl}/\tau_{\rm bb}$.

\begin{figure}[h]
\includegraphics[3.5cm,0.8cm][6.2cm,6.5cm]{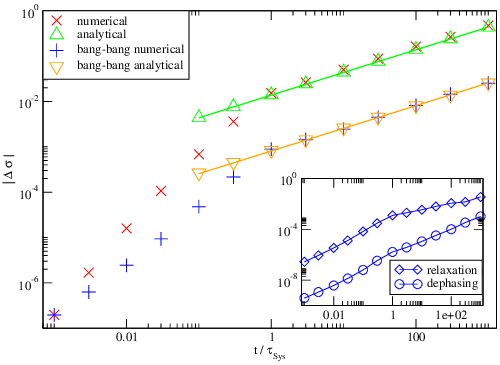}
\caption{\label{devevolution}Time evolution of the mean deviations
for bfl-induced random walks with and without bang-bang. The
straight lines are square-root fits of the analytical derived
random walk model variances (plotted as triangles). Inset:
Transverse and perpendicular components of bang-bang suppressed
noise.}
\end{figure}

As generic conditions of the system dynamics we consider for the
numerical simulations $\epsilon_{\rm q}=\Delta_{\rm q} \equiv
\Omega_0$. Without loss of generality, we assume $\langle
\hat{\sigma}_{\rm z}^{\rm q} \rangle =+1$ as an initial state. If
there were no noise, the spin would precess on the Bloch sphere
around the rotation axis $\hat{\sigma}_{\rm x}^{\rm
q}+\hat{\sigma}_{\rm z}^{\rm q}$. So we expect for not too large
an interaction strength ($\alpha \ll 1$) a slight deviation of the
individual quantum trajectory from the free evolution case.  We
take $\alpha=0.1$ for our coupling strength. All the following
time and energy measures are given in units of the unperturbed
system Hamiltonian: our time unit is $\tau_{\rm Sys} = 1/\Omega_0$
and our energy unit is $\Delta{\rm E} = \sqrt{\epsilon_{\rm q}^2 +
\Delta_{\rm q}^2}=\sqrt{2}\Omega_0$. Note that in these units, a
period lasts $\pi\tau_{\rm sys}/\sqrt{2}$. We have integrated
Equ.~(\ref{ssesolution}) and averaged over $N=1000$ realizations.
The time scale ratio $\tau_{\rm bfl}/\tau_{\rm bb} = 10$ if not
denoted otherwise. We characterize our results by the
root-mean-square deviation from the unperturbed signal
\begin{equation}
\Delta \vec{\sigma}_{\rm rms}(t) =  \sqrt{ \frac{1}{N} \sum_j \left(
\vec{\sigma}^{\rm q}_j(t)-\vec{\sigma}^{\rm q}_{{\rm noisy},j}(t)
\right)^2 }
\end{equation}
In order to characterize the degree of noise suppression by means
of bang-bang control, we define the suppression factors for a
given time $t_0$
\begin{eqnarray}
\mathcal{S}_{t_0}(\tau_{\rm bfl}/\tau_{\rm bb}) & = &
\frac{\Delta\vec{\sigma}_{\rm rms}^{\rm
bfl}(t_0)}{\Delta\vec{\sigma}_{\rm rms}^{\rm bb}(t_0)}.
\end{eqnarray}
The deviation as a function of time is plotted in
Fig.~(\ref{devevolution}). \\
~~\\

We recognize that the total deviations on intermediate time scales 
are suppressed by a ratio of $\simeq 10$. Detailed analysis shows 
that the tangential and the orthogonal deviation, corresponding 
respectively to dephasing and relaxation, are of the same size for 
the uncompensated case.  In contrast, the bang-bang modulation mostly 
compensates the dephasing-type deviation, as shown in the inset of
Fig.~(\ref{devevolution}).

We now develop analytical random walk models for our system.
Although the random walk on the Bloch sphere is in general
two-dimensional, bang-bang control effectively reduces it to a
one-dimensional model, representing the relevant perpendicular
part. We restrict ourselves to the long-time limit.

For simplicity, we replace the fluctuating number of random walk
steps for a given time $\Delta t$ of noisy evolution by its
expectation value $\Delta t/\tau_{\rm bfl}$ \cite{WeissRW}. This
allows to use the number of random walk steps as time parameter.
We encounter different one-step-distributions, depending on
whether the number of steps is odd or even, corresponding to an
``up'' or ``down'' state of the bfl \footnote{We assume the
bfl being initially in its ``upper'' state. This restriction is
of no relevance for the long time limit.}. The step-size
distribution of the bfl model in our small deviation regime is
given from Poisson statistics
\begin{equation}
\Phi_{\rm odd/even}^{\rm bfl} (x)  =  {e^{\mp x / \beta} \theta(\pm x)
\over \beta}
\end{equation}
with $\beta = \alpha \tau_{\rm bfl}$ as a typical random walk
one-step deviation.  $\tau_{\rm unit}$ is a time unit,
corresponding to a discrete step length $x_{\rm unit} =
\alpha~\tau_{\rm unit}$ of the random walk. $\theta(x)$ denotes
the Heaviside step function. We neglect the correlations between
transverse and perpendicular deviations as they average out in the
long-time limit.

For the bang-bang suppressed random walk, the flipping positions
of the bfl-noise sign in the bang-bang time-slots are
essentially randomly distributed as long as $\tau_{\rm
bb}\ll\tau_{\rm bfl}$. We find a homogenous step-size distribution
between zero deviation and a maximum $\gamma = {\alpha 2\tau_{\rm
bb}\over \sqrt{2}}$,
\begin{eqnarray}
\Phi_{\rm odd/even}^{\rm bb} (x) & = & {\theta(\pm x) \theta\left( \pm
[\gamma - x] \right) \over \gamma}.
\end{eqnarray}
The ubiquitous $1 \over \sqrt{2}$ occurs, because the bang-bang sequence
also averages over the static $\epsilon_{\rm q}$-term and hence slows down
the free evolution.

By means of these one-step probability distributions, we are able
to calculate via convolution the distributions for $2N$-step
random walks. Specifically, they are the inverse Fourier
transforms of the $N$-fold products of the Fourier transforms of
the two-step distribution \cite{WeissRW}. For the uncompensated
case, we find
\begin{equation}
\Phi_{\rm 2N}^{\rm bfl}(x) =
\int_{-\pi}^{\pi} \frac{dk}{2\pi\beta^{2N}}  e^{-i k x}\left(
\frac{1}{1-2\cos(k) e^{-1/\beta} + e^{-2/\beta}} \right)^N
\end{equation}
whereas for the compensated case
\begin{eqnarray}
\Phi_{\rm 2N}^{\rm bb}(x) & = & \int_{-\pi}^{\pi}
 \frac{dk}{2\pi\gamma^{2N}} e^{-ikx}\left(
\frac{[1-\cos((\gamma+1)k)]^2}{[1-\cos(k)]^2} \right)^N.
\end{eqnarray}
Already for random walk step-numbers in the order of ten, the
resulting distributions are almost Gaussian.
Their standard deviations give the rms deviations of the random
walk models plotted in Fig.~(\ref{devevolution}). As expected,
they grow as a square-root of the number of steps.

The above integrals can be evaluated analytically using the
saddle point approximation. We find variances of
\begin{equation}
\label{sigmabfl} \sigma_{\rm bfl}(N)  =  \sqrt{2N} \alpha
\tau_{\rm bfl}
\end{equation}
for the pure bfl random walk and
\begin{equation}
\label{sigmabb} \sigma_{\rm bb}(N) =
\sqrt{\frac{2N}{3}}\alpha\tau_{\rm bb}
\end{equation}
for the compensated one. In the large-$N$ limit, this model shows
excellent agreement with the simulation.

\begin{figure}[h]
\includegraphics[width=0.49\columnwidth]{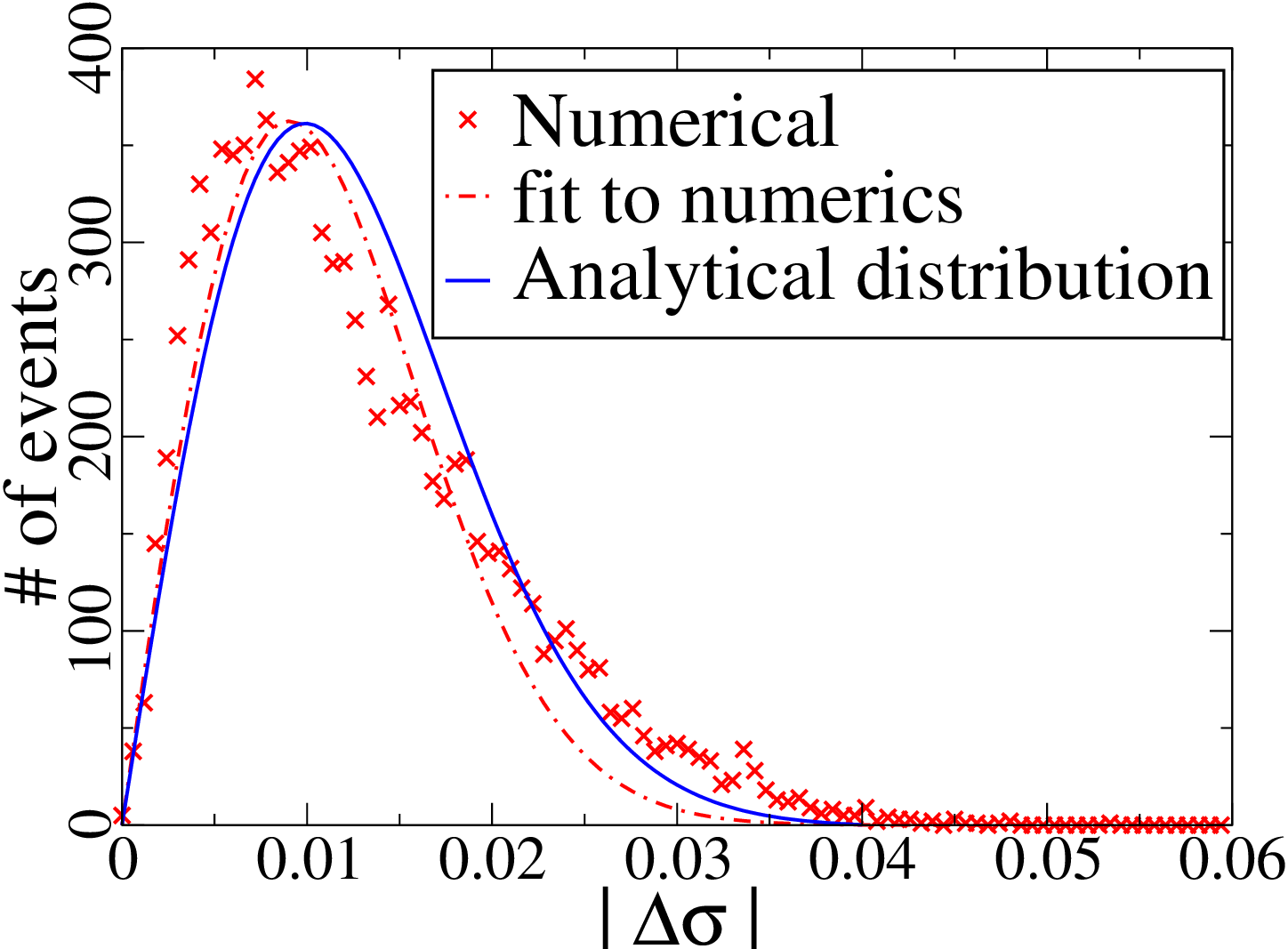}
\includegraphics[width=0.49\columnwidth]{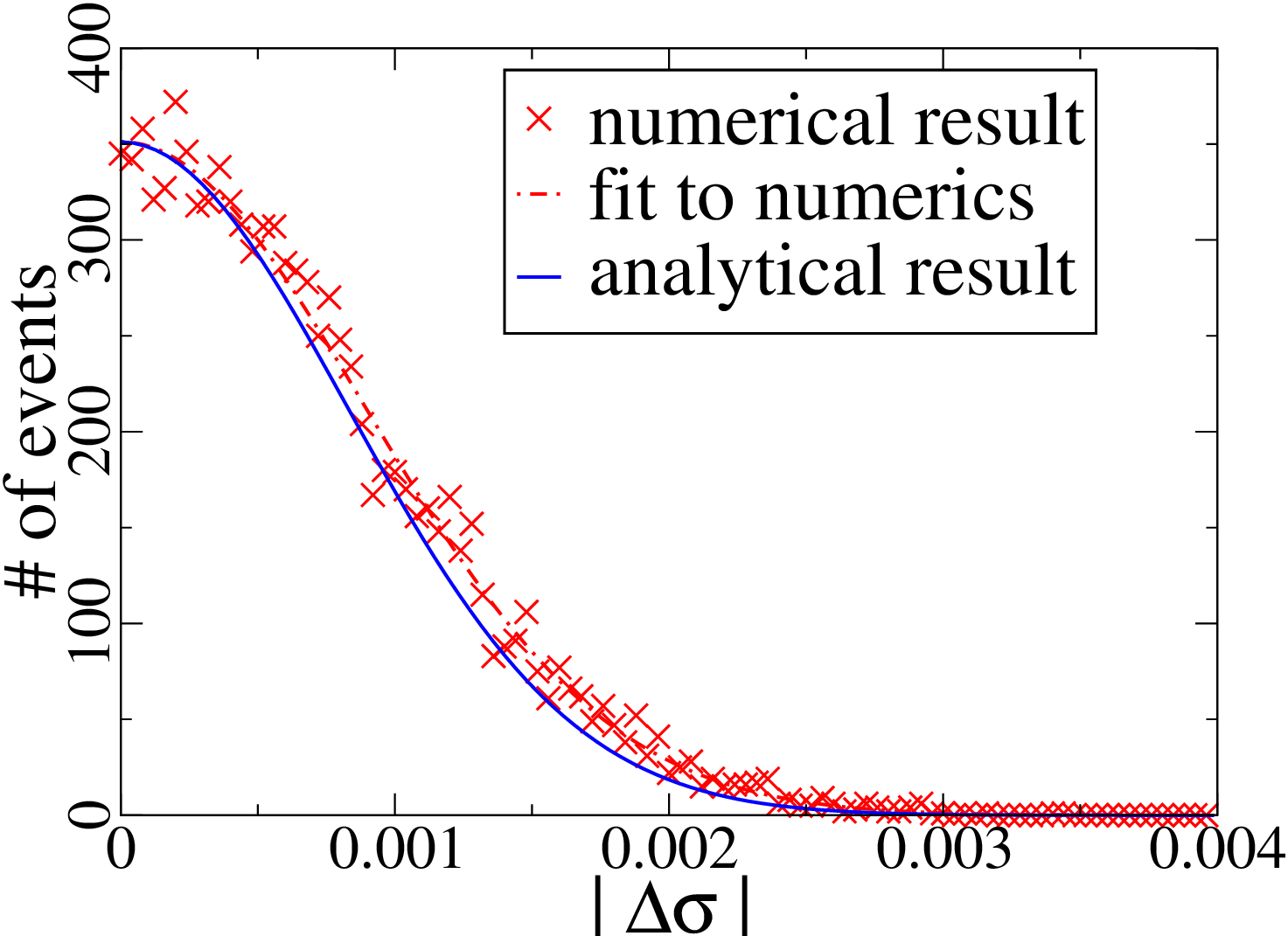}
\caption{\label{distributions}Histograms of the deviation from
free evolution with and without bang-bang and fits to the expected
two- (pure bfl case), respectively one-dimensional (bang-bang
corrected evolution) random walk statistics. Numerical data
collect over 10000 realizations at a fixed time $t_0=\tau_{\rm
Sys}$. With $\tau_{\rm bfl}= 0.01 \tau_{\rm Sys}$ the random walk
distributions are calculated for $N=\tau_{\rm Sys}/\tau_{\rm
bfl}=100$ steps. (NB: The $x$-axis scale of the right graph is 15
times smaller than that of the left graph.)}
\end{figure}

Beyond predicting the variance, our analysis also allows
evaluation of the full distribution. We compared evolution with
and without bang-bang compensation via simulations with $10^4$
realizations and calculated the full distribution function for a
evolution time $t_0=\tau_{\rm Sys}$. The numerical
histograms of the deviation with their respective one- and
two-dimensional Gaussian fits are shown in
Fig.(\ref{distributions}).

We observe that not only the bang-bang compensated distribution is
much narrower than the uncompensated distribution, but also that
its shape is qualitatively different: its maximum is at zero error
whereas the uncompensated distribution has its maximum at a finite
error $|\Delta \sigma| \approx 0.01$ and zero probability of zero
error.

We have systematically studied suppression factors for different
ratios of the switching time $\tau_{\rm bfl}/\tau_{\rm bb}$ at a
constant fluctuator flipping rate $\tau_{\rm bfl}=10^{-2}\tau_{\rm
sys}$ and evolution time $t_0=\tau_{\rm sys}$. The numerical data
in Fig.~(3) show that the suppression efficiency is linear in the
bang-bang repetition rate,  $S=\mu \tau_{\rm bfl}/\tau_{\rm
bb}$. The numerically derived value of the coefficient,
$\mu_{\rm numerical}\approx 1.679$, is in excellent
agreement with our analytical result $\mu_{\rm
analytical}=\sqrt{3}\simeq 1.732$ from the saddle point
approximation, Equs.~(\ref{sigmabfl}) and (\ref{sigmabb}). This
small discrepancy reflects the correlations between the transverse
and longitudinal random walk in the uncompensated case, see
Fig.~(\ref{distributions}).
\begin{figure}[h]
\includegraphics[width=6.6cm]{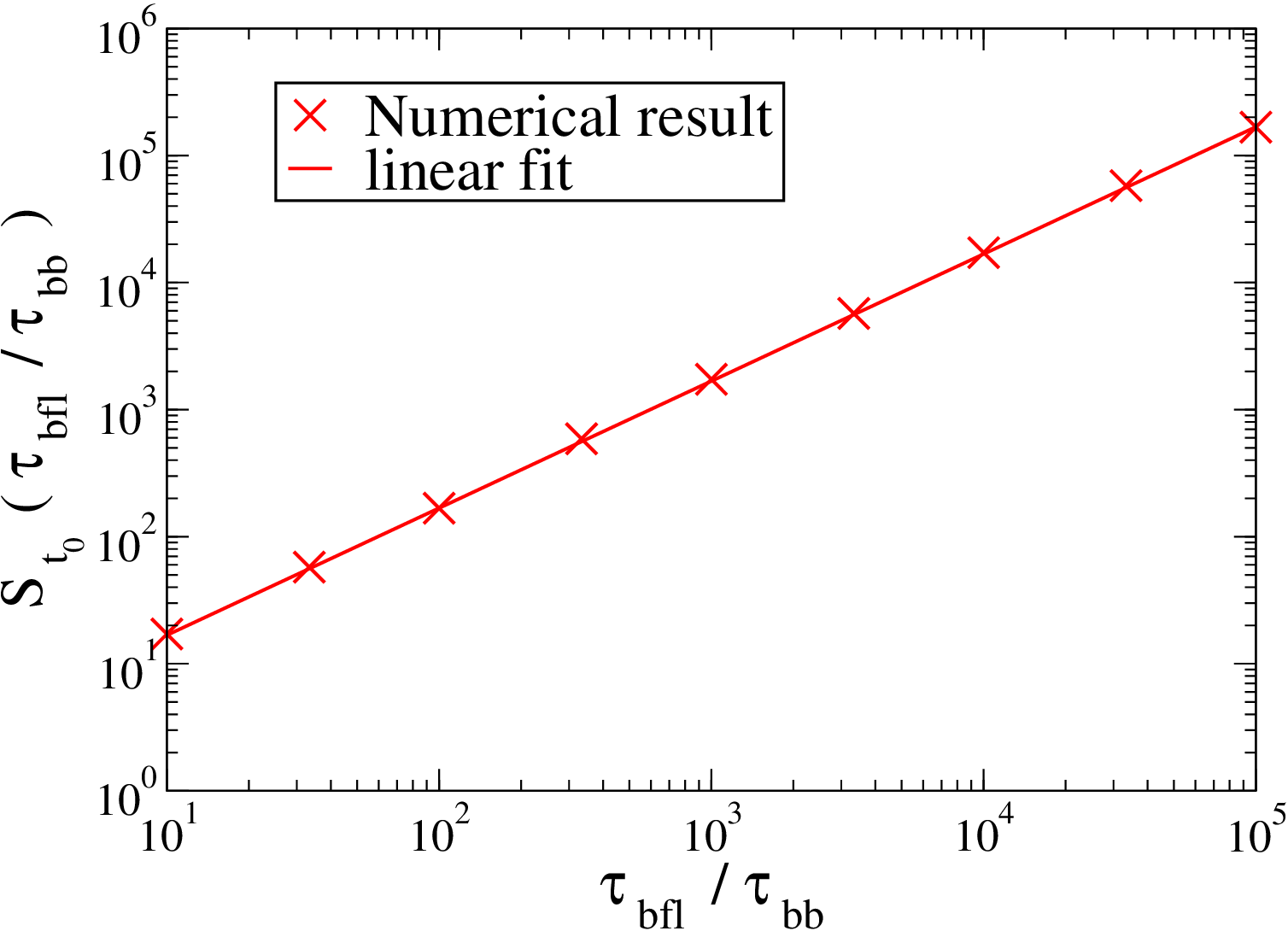}
\caption{The suppression factor $\mathcal{S}_{t_0}(\tau_{\rm
bfl}/\tau_{\rm bb}) = \frac{\Delta\vec{\sigma}_{\rm rms}^{\rm
bfl}(t_0)}{\Delta\vec{\sigma}_{\rm rms}^{\rm bb}(t_0)}$ evaluated
for $t_0=\tau_{\rm Sys}$ as a function of the ratio of the
flipping time $\tau_{\rm bfl}$ and the bang-bang pulse separation
$\tau_{\rm bb}$. \label{fig:suppression}}
\end{figure}

We have demonstrated the ability of a bang-bang protocol to
compensate environmental fluctuations with frequency $\omega\ll
1/\tau_{\rm bb}$. Thus, bang-bang is acting as a ``high pass
filter'' for noise with a roll-off frequency of $1/\tau_{\rm bb}$.
Evidently, the bang-bang correction is suitable for suppressing
the impact of telegraph noise on qubits and can enhance the
coherence by orders of magnitude. The application of the scheme
which we outlined requires a relatively strict separation of time
scales: One has to be able to flip the spin very rapidly,
typically two orders of magnitude faster than $\tau_{\rm bfl}$. It
remains to be investigated how this scheme works with pulse
durations that are finite rather than infinitesimal. Moreover, we
have assumed that the environment produces symmetric telegraph
noise regardless of the qubit dynamics.  Clearly, the issue of
when one may neglect feedback effects between the qubit and bfl
must be critically revisited in the low-temperature limit. We
speculate that the setup is promising for $1/f$-noise, as in
particular the most harmful and predominantly low-frequency
fraction of a corresponding ensemble of fluctuators would be
compensated most strongly. Finally, one has to be aware that also
the static term of the Hamiltonian is averaged out, and this
generally reduces the degree of control on the qubit. This is only
a technical constraint, however, as one could imagine
interchanging two different types of bang-bang pulses to admit
corresponding quantum gate operations.

Another approach for decoupling from slow noise is to choose an
appropriate working point with a dominant term $\Omega\sigma_x$ in
the static Hamiltonian. The action of this term can be understood
as a rapid flipping of the spin, similar to our bang-bang
protocol. Using a Gaussian approximation the noise from the bfl
with standard rate expressions (e.g.,\ \cite{EPJB}), it can be
shown that the dephasing rate reads $\Gamma_\phi=\alpha/\tau_{\rm
bfl}\Omega^2$ instead of $\Gamma_\phi=\alpha\tau_{\rm bfl}$, which
corresponds to the same amount of reduction as in our case. This
scheme has been implemented in superconducting qubits \cite{Vion}.
In that case, it turned out that because the $\sigma_x$ term was
limited by fabrication, this consideration led to a major
redesign. Our compensation scheme purely relies on external
control and thus keeps the hardware design flexible.

A related problem has been addressed in Ref.~\cite{Lidar}, which
deals with bang-bang suppression of {\em Gaussian} $1/f$-noise,
i.e., a bosonic bath with an appropriate sub-Ohmic spectrum. That
system is treated in the weak-coupling approximation, i.e.\ it
assumes $S(\omega)/\omega\ll1$ at low frequencies where
$S(\omega)$ is the noise spectral function. Both assumptions are
serious constraints in the $1/f$-case \cite{PFFF02,D&H}. Our
work is not constrained to weak coupling, takes the full
non-Gaussian statistics of telegraph noise into account, and 
gives the full resulting distribution of errors.

In summary, we examined the decoherence of a single qubit from a
single symmetric telegraph noise source and proposed an adequate
open quantum control compensation protocol for suppressing its
impact. We simulated the qubits dynamics using a stochastic
Schr\"odinger equation and analyzed its deviation from free
evolution. We formulated analytically solvable one- and
two-dimensional random walk models, which are in excellent
agreement with the simulations in the long time limit.
Specifically, we show quantitatively, how the degree of noise
compensation is controlled by the ratio between bfl flipping
time scale and bang-bang pulse length. We give the full statistics
of deviations in both cases.

We thank T.P. Orlando, I. Goychuk, J. von Delft and especially A.
K\"ack for helpful and insights deepening discussions. HG and FKW
are also indebted to T.P. Orlando for his great hospitality at
MIT. WMK gratefully acknowledges fellowship support from the
Fannie and John Hertz Foundation. This work was supported by a
DAAD NSF travel grant, by ARO project P-43385-PH-QC and the DFG
through SFB 631.


\end{document}